\UseRawInputEncoding 
\documentclass[twocolumn,showpacs,superscriptaddress,amsmath,amssymb,aps,pra]{revtex4-1}
\usepackage{graphicx}
\usepackage{dcolumn}
\usepackage{amsmath,txfonts}
\usepackage{epstopdf}
\usepackage[mathlines]{lineno}
\usepackage{color}
\usepackage[colorlinks=true,linkcolor=blue,citecolor=blue]{hyperref}
\usepackage[normalem]{ulem}

\begin{document}

\title{Enhancement of microwave squeezing via parametric down-conversion in a superconducting quantum circuit}

\author{Kong Han}
\affiliation{Interdisciplinary Center of Quantum Information and Zhejiang Province Key Laboratory of Quantum Technology and Device, Department of Physics and State Key Laboratory of Modern Optical Instrumentation, Zhejiang University, Hangzhou 310027, China}

\author{Yimin Wang}
\affiliation{Communications Engineering College, Army Engineering University, Nanjing 210007, China}

\author{Guo-Qiang Zhang}
\email{zhangguoqiang3@zju.edu.cn}
\affiliation{Interdisciplinary Center of Quantum Information and Zhejiang Province Key Laboratory of Quantum Technology and Device, Department of Physics and State Key Laboratory of Modern Optical Instrumentation, Zhejiang University, Hangzhou 310027, China}

\begin{abstract}
We propose an experimentally accessible superconducting quantum circuit, consisting of two coplanar waveguide resonators (CWRs), to enhance the microwave squeezing via parametric down-conversion (PDC). In our scheme, the two CWRs are nonlinearly coupled through a superconducting quantum interference device embedded in one of the CWRs. This is equivalent to replacing the transmission line in a flux-driven Josephson parametric amplifier (JPA) by a CWR, which makes it possible to drive the JPA by a quantized microwave field. Owing to this design, the PDC coefficient can be considerably increased to be about tens of megahertz, satisfying the strong-coupling condition. Using the Heisenberg-Langevin approach, we numerically show the enhancement of the microwave squeezing in our scheme. In contrast to the JPA, our proposed system becomes stable around the critical point and can generate stronger transient squeezing. In addition, the strong-coupling PDC can be used to engineer the photon blockade.
\end{abstract}

\date{\today}

\maketitle

\section{Introduction}

Squeezing, i.e., the reduction of quantum fluctuations in one quadrature component at the expense of increasing fluctuations in the other canonically conjugate variable, is one of the extraordinary effects in quantum optics~\cite{Walls83,Slusher85,Wu86}. Owing to the promising applications in, e.g., precision measurement~\cite{Otterstrom14,Kruse2016,Malnou19}, quantum key distribution~\cite{Garcia-Patron09,WangPu19}, engineering matter interactions~\cite{Lv15,Zeytinoglu17,Qin18} and improving the efficiency of heat engines~\cite{Klaers17,Wang19}, the generation of squeezed light has attracted much attention. Up to now, numerous schemes for squeezing have been proposed, typically using the optomechanical interaction~\cite{Fabre94,Xiao14,Yu18}, the four-wave mixing~\cite{Slusher85}, and the degenerate parametric down-conversion (PDC)~\cite{Wu86}. Here, the degenerate PDC involves a three-wave mixing process of generating a lower-energy photon pair with the {\it same} frequency by splitting a higher-energy photon via a nonlinear medium~\cite{Walls94}.

On the other hand, fast developments of the superconducting quantum circuits (SQCs) have greatly facilitated the progress in quantum information processing (see. e.g., \cite{Devoret1169,You11,Wendin_2017,Tsai_2021}). Since the SQCs based on Josephson junctions can be designed and fabricated to tailor their characteristics for various purposes, they have also been used as the platforms for exploring quantum-optics phenomena in the microwave domain~\cite{You11}. For instance, electromagnetically induced transparency and Autler-Townes splitting~\cite{Li15,Novikov16,Long18,Chien19}, sideband transitions~\cite{Diaz16,Chen2017},quantum entanglement~\cite{Wei2006,Neeley2010,White2016,Armata17,Izmalkov04,Zippilli15}, Bloch-Siegert shifts~\cite{Diaz10,Wang20} and superradiant quantum phase transitions~\cite{Feng15,Xiang16,Motoaki16,Garziano14} were demonstrated using the SQCs. Also, the generation of squeezed microwaves via, e.g., degenerate four-wave mixing and degenerate PDC was proposed theoretically and later realized experimentally~\cite{Yurke89,Cao11,Benlloch14,Zhong_2013,Moon05,Wang15}. Most importantly, the advantage of easy tunability in SQCs makes it promising to demonstrate and manipulate more quantum-optics phenomena in the microwave domain.

In this paper, we propose an experimentally feasible scheme to obtain squeezed microwaves via quantized-field-driven PDC in a SQC. The considered system is composed of two superconducting coplanar waveguide resonators (CWRs), denoted as CWR-A and CWR-B, which are inductively coupled to each other by a superconducting quantum interference device (SQUID) embedded in CWR-A. The magnetic flux threading the SQUID loop contains the externally applied magnetic flux and the {\it quantized} magnetic flux generated by the current in CWR-B. This is equivalent to replacing the transmission line in a flux-driven Josephson parametric amplifier (JPA) with a CWR (i.e., the CWR-B), which makes it possible to drive the JPA by a quantized microwave field. A similar method was also used in the optical parametric amplifier to enhance the squeezing by embedding the nonlinear crystal in an optical cavity~\cite{Vahlbruch16},
where the nonlinear crystal and the optical cavity play similar roles as the nonlinear CWR-A (i.e., the JPA) and the CWR-B in our proposed system, respectively. The generated quantized magnetic flux modulates the phase across the SQUID and gives rise to the mutual coupling between the two resonators. The SQUID usually works in the phase regime where the phase degree of freedom dominates. When the external magnetic field is much stronger than the generated quantized magnetic field, the coupled SQC system can be described by a PDC Hamiltonian, where the classical drive field in the JPA is replaced by a quantized drive field. With this quantized-field-driven PDC Hamiltonian, we study the microwave squeezing in the proposed system. In fact, due to the nonlinearity of the Josephson junction, a small SQUID is a highly nonlinear system. Therefore, we can use it to explore the high-order nonlinear effect of the SQUID on the microwave squeezing in our SQC system.

Around a critical drive strength, our proposed system is stable and can produce optimum steady-state squeezing. This is different from a flux-driven JPA~\cite{Zhong_2013}, which can have the same degree of steady-state squeezing, but is unstable around the critical point, yielding the optimum steady-state squeezing experimentally inaccessible. Moreover, for the JPA, its transient squeezing cannot exceed the steady-state squeezing, while our scheme does not have this limitation and stronger transient squeezing can be achieved (cf. Sec.~\ref{squeezing}). Also, compared with the microwave squeezing in Refs.~\cite{Moon05,Wang15}, our proposal has distinct advantages. First, the design we present here is non-dispersive and directly interacted. This is in contrast to other proposals based on dispersive and indirect couplings via couplers~\cite{Moon05,Wang15}. While in those works the nonlinear coupling strength (which is only tens of kilohertz) is much limited by the dispersive condition, our setup can increase the nonlinear coupling strength by {\it three} orders of magnitude. This promises an appreciable enhancement of the microwave squeezing in the proposed circuit. Second, our setup with the SQUID can be easily fabricated as well, since the needed technology has been maturely utilized for other experiments~\cite{Baust15}. In addition, our scheme may have other potential applications in quantum technologies. For example, the nonlinear coupling strength in our scheme is larger than the decay rates of CWRs (usually smaller than one megahertz~\cite{Xiang13}), so a strong coupling can be realized between the two CWRs. Contrary to the weak-coupling regime, the photon blockade becomes accessible in this strong-coupling regime~\cite{Zhou2020}.

\begin{figure}[tbp]
\centering
\includegraphics[width=0.45\textwidth]{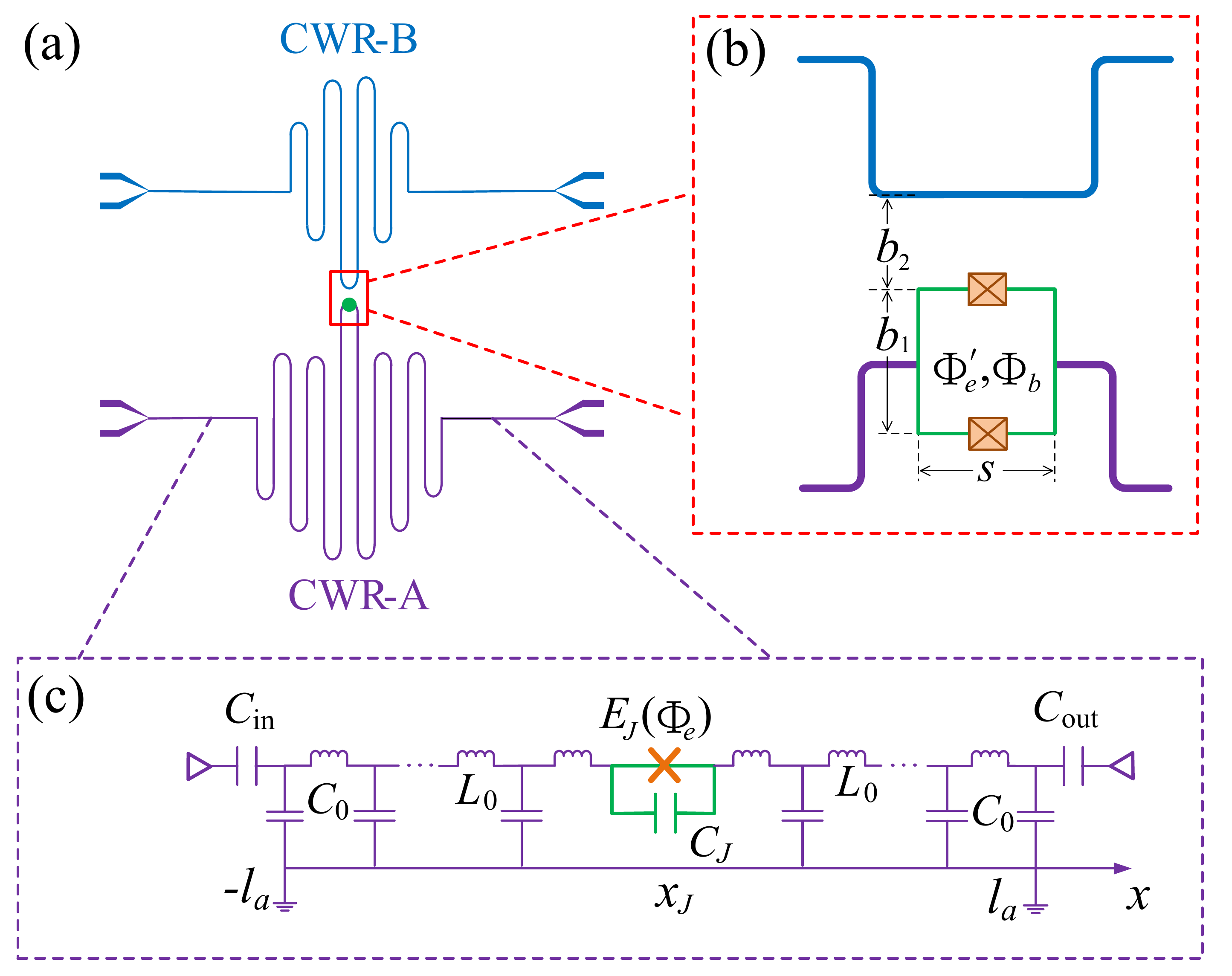}
\caption{(a) Schematic of the proposed SQC system consisting of two CWRs denoted as CWR-A and CWR-B, which are nonlinearly coupled to each other via a SQUID (green dot) embedded in CWR-A. (b) An enlarged view of the part denoted by a red rectangular in (a). (c) The lumped-element circuit of CWR-A, where the SQUID is located at $x=x_{J}$.}
\label{figure1}
\end{figure}

\section{Proposed circuit}\label{model}

As illustrated in Fig.~\ref{figure1}(a), the proposed SQC system is composed of two CWRs (i.e., CWR-A and CWR-B), where both resonators are inductively coupled to each other via a SQUID embedded in CWR-A. Below we first derive CWR-A's Hamiltonian and then give the Hamiltonian of the proposed SQC system.

\subsection{Nonlinear resonator}\label{TLRA}

Due to the SQUID embedded, CWR-A is a nonlinear resonator, which can act as a JPA in the regime with appropriate parameters~\cite{Zhong_2013}. We assume CWR-A to be of a length $2l_{a}$ and decompose it into lumped-circuit elements, as shown in Fig.~\ref{figure1}(c). For simplicity and without loss of generality, we also assume that the two resonators have the same capacitance $C_{0}$ per unit length and the same inductance $L_{0}$ per unit length. When the SQUID is excluded, the Lagrangian of the bare resonator for CWR-A can be written as
\begin{eqnarray}\label{bare}
\mathcal{L}'_{a}=\int_{-l_{a}}^{l_{a}}\bigg[\frac{C_{0}}{2}\dot{\Phi}^{2}(x,t)-\frac{[\partial_{x}\Phi(x,t)]^{2}}{2L_{0}}\bigg]dx,
\end{eqnarray}
where $\Phi(x,t)=\int_{-\infty}^{t}V(x,t')dt'$ is the magnetic flux and $V(x,t')$ is the voltage. For a symmetric SQUID with $C_{J1}=C_{J2} =C_{J}/2$ and $E_{J1}=E_{J2}=E_{J}/2$, the effective Josephson energy is $E_{J}(\Phi_{e})=E_{J}\cos(\pi\Phi_{e}/\Phi_{0})$, where $\Phi_{0}=h/(2e)$ is the flux quantum. The total magnetic flux $\Phi_{e}$ threading the SQUID loop is composed of the externally applied magnetic flux $\Phi'_{e}$ and the magnetic flux $\Phi_{b}$ generated by the current in CWR-B, i.e., $\Phi_{e}=\Phi'_{e}+\Phi_{b}$, where the small magnetic flux induced by the self inductance of the SQUID loop is neglected. Under the condition $\Phi'_{e} \gg \Phi_{b}$, the change of the effective Josephson energy $E_{J}(\Phi_{e})$ due to the magnetic flux $\Phi_{b}$ can be linearly approximated as
$E_{J}(\Phi_{e})\approx E_{J}(\Phi'_{e})[1-\tan(\pi\Phi'_{e}/\Phi_{0})(\pi\Phi_{b}/\Phi_{0})]$. This approximation is reasonable because $0.25\leq\Phi'_{e}/\Phi_{0} \leq0.45$ and $\Phi_{b}/\Phi_{0} \sim 1\times 10^{-3}$ in our paper.

The Lagrangian of the SQUID takes the standard form
$\mathcal{L}_{S}=C_{J}\dot{\delta}^{2}/2- E_{J}(\Phi_{e})\big[1-\cos(2\pi \delta/\Phi_{0})\big]$,
where $\delta=\Phi(x_{J}^{+},t)-\Phi(x_{J}^{-},t)$ is the phase drop of the SQUID located at $x=x_{J}$ [see Fig.~\ref{figure1}(c)]. Another important parameter of the SQUID is the effective charge energy $E_{C}=(2e)^{2}/(2C_{J})$. When the SQUID is in the phase regime, i.e., $E_{J}(\Phi_{e})\gg E_{C}$, we expand the cosine potential of the SQUID up to the fourth order of $2\pi\delta/\Phi_{0}$. The Lagrangian $\mathcal{L}_{S}$ can be approximately written as $\mathcal{L}_{S}=\mathcal{L}_{S}^{(2)}+\mathcal{L}_{S}^{(3)}+\mathcal{L}_{S}^{(4)}$, where
\begin{align}\label{L_S1}
\mathcal{L}_{S}^{(2)}&=\frac{C_{J}}{2}\dot{\delta}^{2}- \frac{\delta^{2}}{2L_{J}}, \notag \\
\mathcal{L}_{S}^{(3)}&=\frac{\delta^{2}}{2L_{J}}\bigg(\pi\frac{\Phi_{b}}{\Phi_{0}}\bigg)\tan\bigg(\frac{\pi\Phi'_{e}}{\Phi_{0}}\bigg), \notag \\
\mathcal{L}_{S}^{(4)}&=\frac{\delta^{4}}{24L_{J}}\bigg(\frac{2\pi}{\Phi_{0}}\bigg)^{2},
\end{align}
with the Josephson inductance $L_{J}$ defined by $1/L_{J} = (2\pi/\Phi_{0})^{2}E_{J}(\Phi'_{e})$. The quality factor of the resonator for CWR-A is assumed to be sufficiently high, so we can consider the limiting case $C_{\rm in,out}\rightarrow 0$, which corresponds to the open-ended boundary condition~\cite{Johansson14}
\begin{eqnarray}\label{boundary1}
\partial_{x}\Phi(x,t)|_{x=-l_{a}}=\partial_{x}\Phi(x,t)|_{x=l_{a}} =0,
\end{eqnarray}
where $C_{\rm in}$ and $C_{\rm out}$ are the capacitances of the input and output capacitors at the two ends of CWR-A.
Moreover, according to Kirchhoff's current law, around $x=x_J$, we have
\begin{eqnarray}\label{boundary2}
\frac{1}{L_{0}}\partial_{x}\Phi(x,t)|_{x=x_{J}^{-}}=\frac{1}{L_{0}}\partial_{x}\Phi(x,t)|_{x=x_{J}^{+}}
=C_{J}\ddot{\delta}+\frac{\delta}{L_{J}},
\end{eqnarray}
due to the presence of the SQUID in the resonator~\cite{Bourassa12}.

First, we study the normal modes of CWR-A and quantize the Hamiltonian related to the linear Lagrangian $\mathcal{L}_{a}=\mathcal{L}'_{a}+\mathcal{L}_{S}^{(2)}$, where higher-order nonlinear terms in the Lagrangian $\mathcal{L}_{S}$ are ignored. As shown in Appendix~\ref{normalmodes}, by using the method of separating variables under the conditions in Eqs.~(\ref{boundary1}) and (\ref{boundary2}), the linear Lagrangian $\mathcal{L}_{a}$ of CWR-A can be written as $\mathcal{L}_{a}=\sum_{m}[C_{\Sigma}\dot{\phi}_{m}^{2}/2-\phi_{m}^{2}/2L_{m}]$. Here $C_{\Sigma}\equiv  2C_{0}l_{a} + C_{J}$ is the total capacitance and $L_m$ is the effective inductance related to the $m$th mode of CWR-A. Both the flux amplitude $\phi_{m}$ and the spatial mode function $\mu_{m}(x)$ of the $m$th mode satisfy $\Phi(x,t)=\sum_{m}\phi_{m}(t)\mu_{m}(x)$. With a Legendre transformation, the Hamiltonian $H_{a}$ of CWR-A is obtained as
\begin{eqnarray}\label{H_a}
H_{a}=\sum_{m}\bigg[\frac{p_{m}^{2}}{2C_{\Sigma}}+\frac{\phi_{m}^{2}}{2L_m}\bigg],
\end{eqnarray}
where $p_{m}=C_{\Sigma}\dot{\phi}_{m}$ is the canonical momentum of the $m$th mode, conjugate to the canonical coordinate $\phi_{m}$. We treat the canonical variables $\phi_{m}$ and $p_{m}$ as quantum operators satisfying the commutation relation $[\phi_{n},p_{m}]=i\hbar\delta_{nm}$ and can then write them as
\begin{align}\label{phip_m}
\phi_{m}&=\sqrt{\frac{\hbar}{2C_{\Sigma}{\omega_{m}}}}(a^{\dag}_{m}+a_{m}),\notag\\
p_{m}&=i\sqrt{\frac{\hbar C_{\Sigma}{\omega_{m}}}{2}}(a^{\dag}_{m}-a_{m}),
\end{align}
where $a_{m}^{\dag}$ and $a_{m}$ are the creation and annihilation operators of the $m$th resonator mode and $\omega_m=1/\sqrt{C_{\Sigma}L_m}$ is the resonant frequency of this mode. If we only focus on the fundamental mode of the resonator (i.e., $m=1$), the Hamiltonian (\ref{H_a}) when setting $\hbar=1$ is reduced to $H_{a}=\omega_{a}a^{\dag}a$, with $a\equiv a_1$ and $\omega_a\equiv\omega_1$. The resonant frequency $\omega_a$ satisfies $\omega_a=k_1\upsilon$, where $k_1$ is the wave vector of the fundamental mode and $\upsilon=1/\sqrt{C_0L_0}$ is the group velocity.

\subsection{Hamiltonian of the proposed system}\label{TLRA-B}

We consider CWR-B to be a $\lambda/2$-type resonator and focus on its fundamental mode as well, i.e., the $\lambda/2$ mode. The Hamiltonian of this mode can be written as $H_{b}=\omega_{b}b^{\dag}b$, with $\omega_{b}=\pi\upsilon/(2l_{b})$ and $b^{\dag}$ ($b$) being the resonant frequency and the creation (annihilation) operator of the mode, respectively. Here, $2l_b$ is the length of the resonator. The quantized current in the waveguide of CWR-B reaches its maximum $I=\sqrt{\hbar{\omega_{b}}/(2L_{0}l_{b})}(b^{\dag}+b)$ at the antinode~\cite{Sun06}. Correspondingly, the magnetic field generated by this current reads $B=\mu_{0}I/(2\pi r)$, where $\mu_{0}$ is the vacuum permeability and $r$ is the radial distance away from the waveguide of CWR-B. The SQUID loop can be designed as a rectangular loop with length $s$ and width $b_{1}$, and it has a distance $b_{2}$ away from the waveguide of the resonator CWR-B [see  Fig.~\ref{figure1}(b)]. Therefore, in the SQUID loop, the part of the quantized magnetic flux generated by the current in CWR-B can be written as
\begin{eqnarray}\label{Phi_b}
\Phi_{b}=\int_{S}B~dS=\phi_{b}(b^{\dag}+b),
\end{eqnarray}
which drives the JPA (i.e., the CWR-A) to produce microwave squeezing (cf. Sec.~\ref{squeezing}), where
\begin{eqnarray}
\phi_{b}=\frac{\mu_{0}s}{2\pi}\bigg(\frac{\hbar{\omega_{b}}}{2L_{0}l_{b}}\bigg)^{1/2}\ln\bigg(\frac{b_{1}+b_{2}}{b_{2}}\bigg).
\end{eqnarray}
This differs from the conventional JPA driven by a classical field~\cite{Zhong_2013}.

We then substitute Eqs.~(\ref{phip_m}) and~(\ref{Phi_b}) into the coupling term $\mathcal{L}_{S}^{(3)}$ given in Eq.~(\ref{L_S1}) with $\delta=\Delta_{1}\phi_{1}$, where $\Delta_1=\mu_{1}(x_{J}^{+})-\mu_{1}(x_{J}^{-})$ is the fundamental-mode amplitude difference across the SQUID (see Appendix~\ref{normalmodes}). The interaction Hamiltonian $H_{I}=-\mathcal{L}_{S}^{(3)}$ is then given by
\begin{equation}\label{HI}
H_I=-\chi(a^{\dag}+a)^{2}(b^{\dag}+b),
\end{equation}
where the nonlinear coupling strength $\chi$ depends on $\Phi'_{e}$,
\begin{eqnarray}
\chi=\tan\left(\pi\frac{\Phi'_{e}}{\Phi_{0}}\right)\frac{\phi_{b}}{\Phi_{0}}\frac{\pi\hbar\Delta_{1}^{2}}
{4L_{J}C_{\Sigma}\omega_{a}}.
\end{eqnarray}

In the case of $\chi \ll \{\omega_{a},\omega_{b}\}$ and $\omega_{b} \approx 2\omega_{a}$, we can neglect the fast oscillating terms via the rotating-wave approximation (RWA)~\cite{Scully97} and the interaction Hamiltonian is reduced to $H_{I}=-\chi(a^{\dag}a^{\dag}b+a ab^{\dag})$. Now, we can write the total Hamiltonian $H_{0}=H_{a}+H_{b}+H_{I}$ of the proposed SQC system as
\begin{eqnarray}\label{equ11}
H_{0}=\omega_{a}a^{\dag}a+\omega_{b}b^{\dag}b-\chi(a^{\dag}a^{\dag}b+a ab^{\dag}) ,
\end{eqnarray}
which is the Hamiltonian for quantized-field-driven PDC.

In another case when the resonant frequency of CWR-A is much larger than that of CWR-B, i.e., $\omega_{a} \gg \omega_{b}$, $H_{I}$ in Eq.~(\ref{HI}) can be approximated as the standard interaction Hamiltonian of an optomechanical system $H_{I}=-2\chi a^{\dag}a(b^{\dag}+b)$~\cite{Aspelmeyer14}. Here it is not the scope of the present work and will not be investigated. In Ref.~\cite{Johansson14}, the case of $x_{J}/l_{a}=\pm 1$ is studied, where the effect of the SQUID is approximated as a turnable length of the SQUID-terminated CWR and the Hamiltonian of two coupled resonators has the standard optomechanical form.

\begin{figure}[tbp]
\centering
\includegraphics[width=0.48\textwidth]{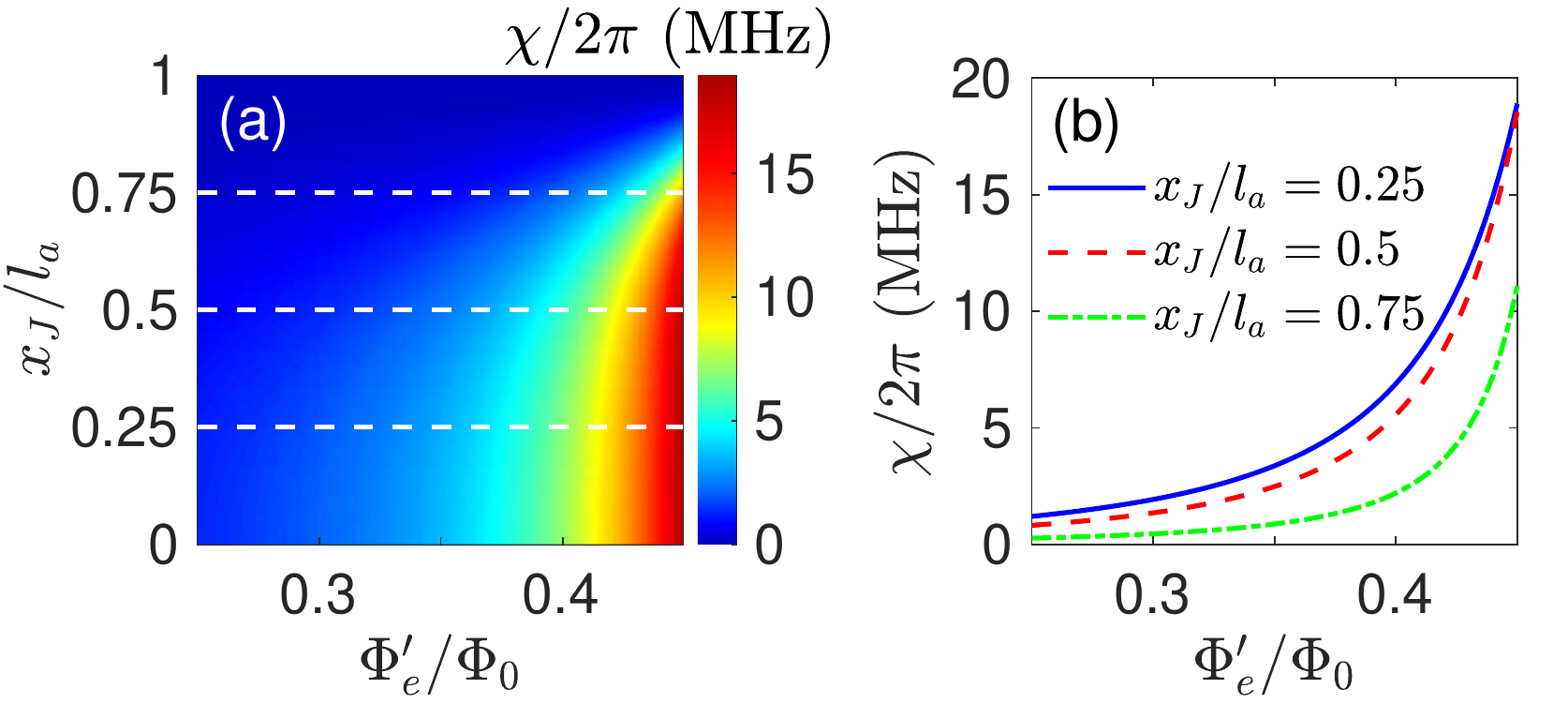}
\caption{(a) The nonlinear coupling strength $\chi/2\pi$ versus the external flux $\Phi'_{e}/\Phi_{0}$ and the position $x_{J}/l_{a}$ of the SQUID. (b) The nonlinear coupling strength $\chi$ versus the external flux $\Phi'_{e}/\Phi_{0}$ for $x_{J}/l_{a}=$ 0.25, 0.5 and 0.75, which are the cross-section views of (a) indicated by the three white dashed lines. Other parameters are $2l_{a}=12$ mm, $2l_{b}=\pi\upsilon/(2\omega_{a})$, $L_{0}=4.5\times 10^{-7}$ H/m, $C_0=1.8\times 10^{-10}$ F/m, $E_{J}/2\pi=600$ GHz, $E_{C}/2\pi=1$ GHz, $s/(2l_{a})=0.05$, and $b_{2}=0.5b_{1}$.}
\label{figure2}
\end{figure}

In Fig.~\ref{figure2}(a), we plot the nonlinear coupling strength $\chi$ versus both the external magnetic flux $\Phi'_{e}/\Phi_{0}$ and the position $x_{J}/l_{a}$ of the SQUID. When the SQUID is located at the position near the center of CWR-A and the reduced external flux $\Phi'_{e}/\Phi_{0}$ approaches 0.45, i.e., around the bottom right corner of Fig.~\ref{figure2}(a), a large coupling strength can be obtained. For clarity, in Fig.~\ref{figure2}(b), we also show the results for $x_{J}/l_{a}= 0.25, 0.5$ and $0.75$, corresponding to the three white dashed lines in Fig.~\ref{figure2}(a). For $x_{J}/l_{a}=0.25$ and $\Phi'_{e}/\Phi_{0}=0.45$, the nonlinear coupling strength is $\chi/2\pi = 18$ MHz, approximately {\it three} orders of magnitude higher than that achieved in previous studies~\cite{Moon05,Wang15}. Since the frequencies of the CWR-A and CWR-B, $\omega_{a}$ and  $\omega_{b}$, are in the microwave domain (a few gigahertz), the condition $\chi \ll \{\omega_{a},\omega_{b}\}$ to make the RWA in Eq.~(\ref{equ11}) is well satisfied. As shown in Sec.~\ref{squeezing}, a larger nonlinear coupling strength can give rise to an appreciable enhancement of the microwave squeezing in the proposed circuit, which reveals the advantage of our system.

The decay rate of a superconducting CWR is usually smaller than $1 \times 2\pi$~MHz~\cite{Xiang13}. Therefore, the proposed SQC system can reach the strong-coupling regime, i.e., $\chi$ is larger than the decay rates of CWR-A and CWR-B, where the photon blockade, inaccessible in the weak-coupling regime, can be engineered via the quantized-field-driven PDC~\cite{Zhou2020}. In fact, under the conditions of both $\Phi'_{e}\gg\Phi_{b}$ and $E_{J}(\Phi_{e}) \gg E_{c}$, an even larger coupling strength $\chi$ can be achieved by further increasing the static magnetic flux $\Phi'_{e}$. Thus, our setup also provides the advantage to demonstrate the photon blockade in a SQC system.

\section{Microwave squeezing enhancement}\label{squeezing}

In this section, we study the performance of our scheme for microwave squeezing. When a microwave field with frequency $\omega_{d}$ drives CWR-B, the interaction Hamiltonian is $H_{d}=\Omega_{d}(b^{\dag}e^{-i\omega_{d}t}+be^{i\omega_{d}t})$, where $\Omega_{d}$ is the drive strength. In the rotating frame with respect to the drive-field frequency $\omega_{d}$, when $\omega_{b}=2\omega_{a}$ and $\omega_{d}=\omega_{b}$, the total Hamiltonian $H=H_{0}+H_{d}$ of the system can be written as
\begin{equation}\label{tot}
H=-\chi(a^{\dag}a^{\dag}b+aab^{\dag})+\Omega_{d}(b^{\dag}+b).
\end{equation}

Intuitively, by taking the parametric approximation (i.e., replacing the operators $b$ and $b^{\dag}$ with their expectation values $\beta$ and $\beta^{*}$) and removing the constant terms, the above Hamiltonian can be approximated as~\cite{Moon05}
\begin{equation}\label{parametric-approximation}
H \approx -(\chi_{\rm eff} a^{\dag}a^{\dag}+\chi_{\rm eff}^{*}aa),
\end{equation}
with $\chi_{\rm eff}=\chi\beta$, which is a standard two-photon Hamiltonian for generating the squeezing. To be more precise, here we do not adopt this approximation, but directly harness the Hamiltonian in Eq.~(\ref{tot}) to study the microwave squeezing of the proposed system.

\begin{figure}[tbp]
\centering
\includegraphics[width=0.48\textwidth]{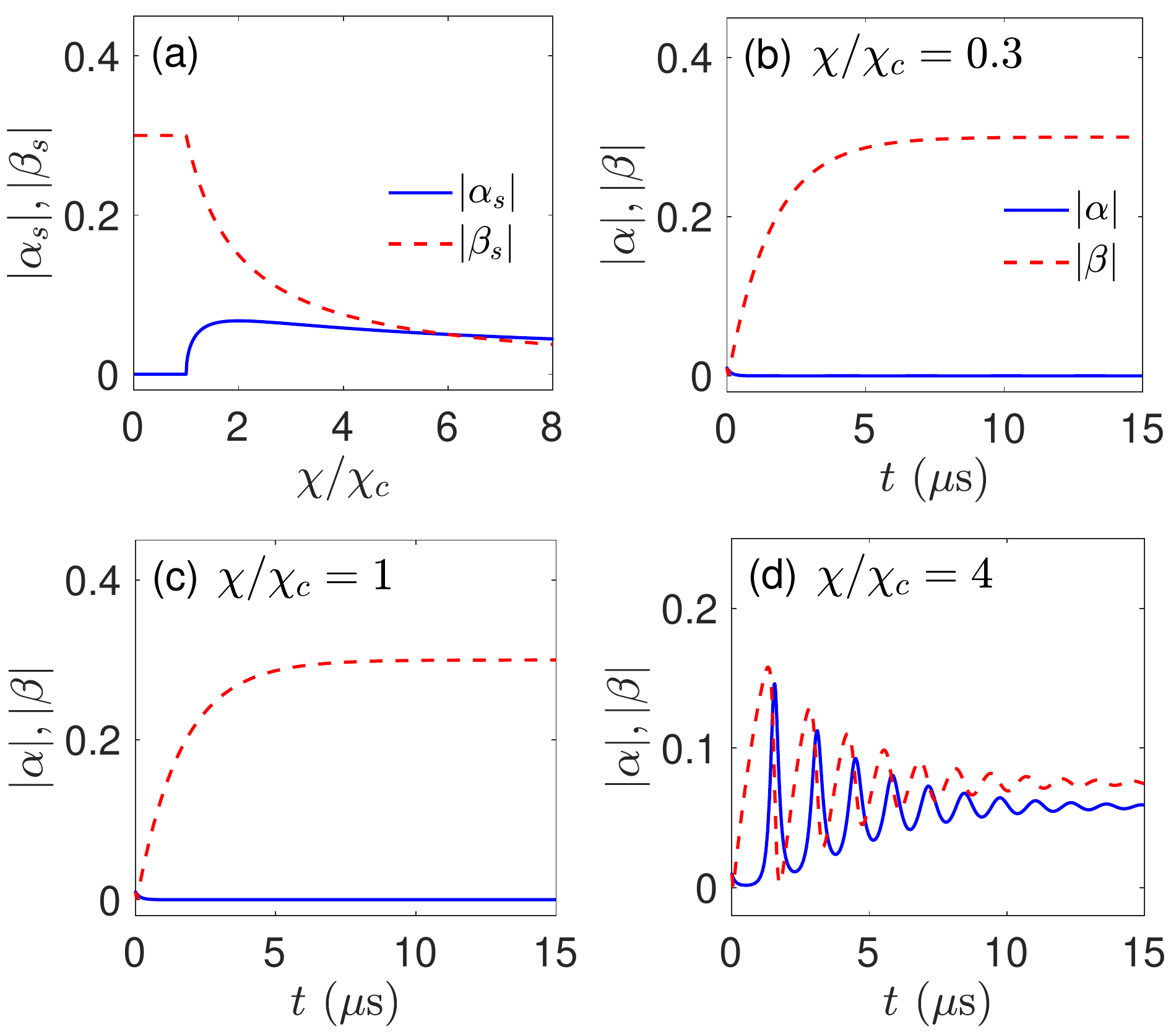}
\caption{(a) The microwave-field amplitudes $|\alpha_s|$ and $|\beta_s|$ versus the reduced coupling strength $\chi/\chi_c$, calculated using Eqs.~(\ref{solution-1}) and (\ref{solution-2}).
(b)-(d) The time evolution of the microwave-field amplitudes $|\alpha|$ and $|\beta|$, calculated using Eq.~(\ref{steady}), where the initial values are $\alpha(t=0)=0.01$ and $\beta(t=0)=0.01i$, with (b) $\chi/\chi_c=0.3$, (c) $\chi/\chi_c=1$, and (d) $\chi/\chi_c=4$. Since the fabrication of the SQUID may reduce the quality factor of CWR-A by one or two orders of magnitude~\cite{Zhong_2013}, we choose $\kappa_{b}=0.1\kappa_{a}$. Other parameters are $\kappa_a/2\pi=2$ MHz and $\Omega_d/2\pi=0.03$ MHz.}
\label{fig-s1}
\end{figure}

We assume that the CWR-A and CWR-B interact with their respective thermal reservoirs, which are independent from each other. Via the Heisenberg-Langevin approach~\cite{Scully97}, the equations of motion for the field operators $a$ and $b$ can be written as
\begin{align}\label{operatorab}
\dot a&=2i\chi a^{\dag}b-\frac{\kappa_{a}}{2}a+F_a(t),\notag\\
\dot b&=i\chi aa-i\Omega_{d}-\frac{\kappa_{b}}{2}b+F_b(t),
\end{align}
where $\kappa_a$ and $\kappa_b$ are the decay rates of CWR-A and CWR-B, respectively, while $F_a(t)$ and $F_b(t)$ are the related noise operators. In order to linearize Eq.~(\ref{operatorab}), we write the operator $a$ $(b)$ as a sum of the expectation value $\alpha$ $(\beta)$ and the fluctuation $\delta a$ $(\delta b)$, i.e., $a=\alpha+\delta a$ and $b=\beta+\delta b$. It follows from Eq.~(\ref{operatorab}) that the expectation values $\alpha$ and $\beta$ satisfy
\begin{align}\label{steady}
\dot \alpha&=2i\chi \alpha^*\beta-\frac{\kappa_{a}}{2}\alpha,\notag\\
\dot \beta&=i\chi \alpha^{2}-i\Omega_{d}-\frac{\kappa_{b}}{2}\beta.
\end{align}

At the steady state, $\dot \alpha=\dot \beta=0$. Solving Eq.~(\ref{steady}) with $\dot \alpha=\dot \beta=0$, we obtain two sets of solutions. One set of solutions are
\begin{align}\label{solution-1}
\alpha_s=0,~~\beta_s=-2i\Omega_{d}/\kappa_{b},
\end{align}
and the other set of solutions are
\begin{align}\label{solution-2}
\alpha_s=\pm\sqrt{\Omega_{d}(\chi-\chi_c)}/\chi,~~\beta_s=-i\kappa_{a}/4\chi,
\end{align}
which are steady for $\chi < \chi_c$ and $\chi > \chi_c$~\cite{Wang15}, respectively, with $\chi_{c}=\kappa_{a}\kappa_{b}/(8\Omega_d)$ being the {\it critical} coupling strength. Due to the occurrence of the second-order phase transition, the microwave-field amplitudes in both CWR-A and CWR-B display abrupt changes at $\chi/\chi_c=1$, as shown in Fig.~\ref{fig-s1}(a).
When $\chi/\chi_c<1$, the absolute values $|\alpha_s|$ and $|\beta_s|$, given by Eq.~(\ref{solution-1}), are constants. However, in the case of $\chi/\chi_c>1$, as the reduced coupling strength $\chi/\chi_{c}$ increases, the absolute value $|\alpha_s|$, given by Eq.~(\ref{solution-2}), increases from 0 at $\chi/\chi_c=1$ to 0.07 around $\chi/\chi_c=2$ and then decreases monotonically (blue solid curve), while the absolute value $|\beta_s|$, also given by Eq.~(\ref{solution-2}), monotonically decreases with the reduced coupling strength $\chi/\chi_c$ (red dashed curve).

To further study the dynamics of the system, we plot the time evolution of the absolute values $|\alpha|$ and $|\beta|$ for various values of the reduced coupling strength $\chi/\chi_{c}$ in Figs.~\ref{fig-s1}(b)-\ref{fig-s1}(d). It can be seen that below the critical coupling strength (i.e., $\chi/\chi_{c} < 1$), the microwave field in CWR-A is almost in the ground state with $|\alpha| \approx 0$ at any time $t$, but the microwave-field amplitude $|\beta|$ of CWR-B increases with the time and finally reaches its steady-state value $|\beta_s|=0.3$ [see Figs.~\ref{fig-s1}(b) and \ref{fig-s1}(c)]. By comparing Figs.~\ref{fig-s1}(b) and \ref{fig-s1}(c), we find that the evolution of $|\alpha|$ ($|\beta|$) remains nearly the same for different values of $\chi/\chi_c$, indicating that in the case of $\chi/\chi_c < 1$, the evolutions are independent of the reduced coupling strength $\chi/\chi_c$ and there is no energy exchange between the two resonators. The reason is that the initial state of CWR-A is nearly in its steady state, while the initial state of CWR-B deviates considerably from its steady state. However, when $\chi$ is larger than the critical value $\chi_{c}$ and the initial states of both CWR-A and CWR-B deviate from their steady states, e.g., $\chi/\chi_{c}$=4 in Fig.~\ref{fig-s1}(d), the time evolutions of $|\alpha|$ and $|\beta|$ exhibit obvious oscillations before reaching their steady-state values $|\alpha_s|=0.06$ and $|\beta_s|=0.08$, and both CWR-A and CWR-B have energy exchange.

\begin{figure}[tbp]
\centering
\includegraphics[width=0.48\textwidth]{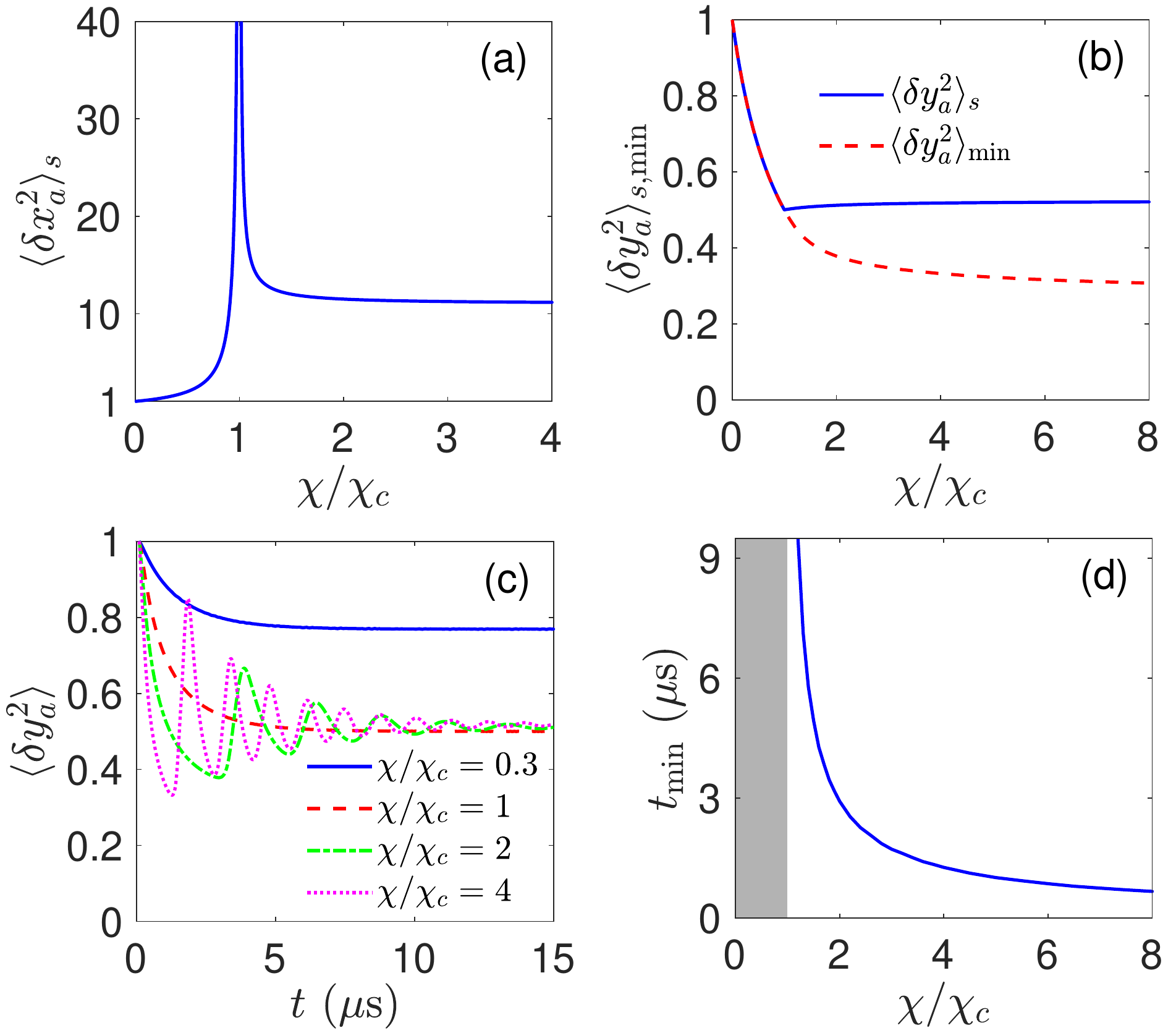}
\caption{(a) The steady-state variance $\langle\delta x_a^2\rangle_s$ versus the reduced coupling strength $\chi/\chi_c$. (b) The steady-state variance $\langle\delta y_a^2\rangle_s$ and the minimum $\langle\delta y_a^2\rangle_{\rm min}$ versus the reduced coupling strength $\chi/\chi_c$. (c) The time evolution of the variance $\langle\delta y_a^2\rangle$ for $\chi/\chi_c=0.3$, 1, 2 and 4. (d) The time $t_{\rm min}$ versus the reduced coupling strength $\chi/\chi_c$. Note that no $t_{\rm min}$ exists in the gray region of $\chi/\chi_{c}<1$. Other parameters are the same as in Fig.~\ref{fig-s1}(a).
}\label{fig-squeezing-a}
\end{figure}

Below we investigate the microwave squeezing in CWR-A via the variances $\langle\delta x_a^2\rangle\equiv\langle x_a^2\rangle-\langle x_a\rangle^2=\langle(\delta a^{\dag}+\delta a)^2\rangle$ and $\langle\delta y_a^2\rangle\equiv\langle y_a^2\rangle-\langle y_a\rangle^2=-\langle(\delta a^{\dag}-\delta a)^2\rangle$, where $x_a=a^{\dag}+a$ and $y_a=i(a^{\dag}-a)$ are the Hermitian amplitude operators. For clarity, here we only give the main results; detailed derivations can be found in Appendix~\ref{dynamic}. At the steady state, the variances $\langle\delta x_a^2\rangle$ and $\langle\delta y_a^2\rangle$ can be expressed as
\begin{eqnarray}\label{variancex1}
\langle\delta x_a^2\rangle_s=\left\{\begin{matrix}
           \dfrac{\kappa_a\kappa_b}{\kappa_a\kappa_b-8\chi\Omega_d},  &  \chi<\chi_c; \\
           1+\dfrac{\kappa_a}{\kappa_b}+\dfrac{\kappa_a\kappa_b}{2(8\chi\Omega_d-\kappa_a\kappa_b)}, &  \chi>\chi_c,
                      \end{matrix}\right.
\end{eqnarray}
and
\begin{eqnarray}\label{variancey1}
\langle\delta y_a^2\rangle_s=\left\{\begin{matrix}
           \dfrac{\kappa_a\kappa_b}{\kappa_a\kappa_b+8\chi\Omega_d},  &  \chi<\chi_c; \\
           \dfrac{16\chi\Omega_d(\kappa_a+\kappa_b)-\kappa_a\kappa_b^2}{16\chi\Omega_d(2\kappa_a+\kappa_b)}, &  \chi>\chi_c.
                      \end{matrix}\right.
\end{eqnarray}

At the critical point $\chi/\chi_c=1$, the variance $\langle\delta y_a^2\rangle_{s}$ has an abrupt change, while the variance $\langle\delta x_a^2\rangle_{s}$ is divergent, which is a characteristic of the second-order phase transition~\cite{Zhu20}. In Figs.~\ref{fig-squeezing-a}(a) and \ref{fig-squeezing-a}(b), we plot the steady-state variances $\langle\delta x_a^2\rangle_s$ and $\langle\delta y_a^2\rangle_s$ versus the reduced coupling strength $\chi/\chi_c$ (blue solid curves). The steady-state variance $\langle\delta x_a^2\rangle_s$ is always larger than 1, while the conjugate variance $\langle\delta y_a^2\rangle_s$ is smaller than 1. This implies the emergence of microwave squeezing in CWR-A~\cite{Wang15,Moon05}. When $\chi/\chi_c<1$, $\langle\delta y_a^2\rangle_s$ decreases monotonically from $1$ to $0.5$, while it increases monotonically from $0.5$ to $0.524$ in the region of $\chi/\chi_c>1$, where the optimum steady-state squeezing is $\langle\delta y_a^2\rangle_{s}=0.5$ around the critical point $\chi/\chi_c=1$.

Moreover, we also present the time evolution of the variance $\langle\delta y_a^2\rangle$ for different values of the reduced coupling strength $\chi/\chi_c$ in Fig.~\ref{fig-squeezing-a}(c). It can be seen that below the critical coupling strength, the variance $\langle\delta y_a^2\rangle$ decreases monotonically with time $t$ to its minimum $\langle\delta y_a^2\rangle_{\rm min}$ ($=\langle\delta y_a^2\rangle_s$), corresponding to the optimum squeezing (cf. the blue solid and red dashed curves). However, when the coupling strength exceeds the critical value $\chi_{c}$, the variance $\langle\delta y_a^2\rangle$ decreases monotonically with time $t$ to its minimum $\langle\delta y_a^2\rangle_{\rm min}$ ($<\langle\delta y_a^2\rangle_s$) and then oscillates before reaching the steady-state value $\langle\delta y_a^2\rangle_s$. These dynamical behaviors of the microwave squeezing can be understood using the parametric-approximation Hamiltonian in Eq.~(\ref{parametric-approximation}). Below the critical coupling strength, the effective nonlinear coefficient $|\chi_{\rm eff}|$ ($\propto |\beta|$) increases monotonically with time $t$ [cf. Figs.~\ref{fig-s1}(b) and \ref{fig-s1}(c)], which is responsible for the monotonic decrease of the variance $\langle\delta y_a^2\rangle$. When $\chi/\chi_{c}>1$ (i.e., above the critical coupling strength), the oscillatory $|\chi_{\rm eff}|$ ($\propto |\beta|$) results in the oscillating behavior of the variance $\langle\delta y_a^2\rangle$ [cf. Fig.~\ref{fig-s1}(d)].

In addition, we plot the minimum $\langle\delta y_a^2\rangle_{\rm min}$ versus the reduced coupling strength $\chi/\chi_c$ in Fig.~\ref{fig-squeezing-a}(b) (the red dashed curve) as well as the time $t_{\rm min}$ (for reaching  $\langle\delta y_a^2\rangle_{\rm min}$) versus $\chi/\chi_c$ in Fig.~\ref{fig-squeezing-a}(d). Note that $\langle\delta y_a^2\rangle$ decreases monotonically with time $t$ in the region of $\chi/\chi_{c}<1$ and the time $t_{\rm min}$ does not exist [see the gray region in Fig.~\ref{fig-squeezing-a}(d)]. Because both $\langle\delta y_a^2\rangle_{\rm min}$ and $t_{\rm min}$ versus $\chi/\chi_c$ decrease monotonically, a larger coupling strength can yield more appreciable and faster squeezing. This promises a considerable enhancement of the microwave squeezing via the quantized-field-driven PDC in our proposed circuit, where the PDC coefficient is strengthened by {\it three} orders of magnitude, from tens of kilohertz in Ref.~\cite{Moon05,Wang15} to tens of megahertz in the present work. When generating the squeezing, the amplitude of the quantized microwave field in CWR-B is on the order of $\kappa_{a}/\chi$ [cf. Eqs.~(\ref{solution-1})-(\ref{solution-2}) and Fig.~\ref{fig-s1}], smaller than 1 because the proposed system is in the strong-coupling regime with $\{\kappa_{a},\kappa_{b}\}_{\rm max} < \chi$. This reveals that microwave squeezing is implemented in the CWR-A, with CWR-B in the quantum limit of single photon.

The above results are obtained at zero temperature, i.e., we take $n_{a}=n_{b}=0$ in the correlation functions $\langle F_q^{\dag}(t)F_q(t')\rangle=n_{q}\kappa_q\delta(t-t')$ and $\langle F_q(t)F_q^{\dag}(t')\rangle=(n_{q}+1)\kappa_q\delta(t-t')$ with $q=a,b$ (cf. Appendix~\ref{dynamic}), where $n_{a}$ and $n_{b}$ are the numbers of thermal average photons in CWR-A and CWR-B. In experiments, the SQCs operate at milli-Kelvin temperatures~\cite{Gu17}, and the typical frequencies of CWRs are a few gigahertz. If we choose $\omega_{a}/2\pi=3.5$~GHz and $\omega_{b}=2\omega_{a}$, the numbers of thermal average photons in CWR-A and CWR-B are about $n_{a}=2.2\times 10^{-4}$ and $n_{b}=5\times 10^{-8}$ at 20 mK ($n_{a}=1.5\times 10^{-2}$ and $n_{b}=2.2\times 10^{-4}$ even at 40~mK), respectively. Therefore, it is reasonable to neglect the temperature effect.

Note that when replacing CWR-B with a transmission line, the system is reduced to a flux-driven JPA, which is a standard device for creating microwave squeezing in the SQCs~\cite{Gu17}. For the JPA driven by a classical field via the transmission line, the Hamiltonian of the system can be written as~\cite{Zhong_2013} $H_{\rm JPA}=i\Omega_d(a^{\dag2}-a^2)$. For simplicity, we use the same symbol $\Omega_{d}$ in both cases. When $\Omega_d/\kappa_{a} < 0.25$, the steady-state value coincides the minimum of the variance $\langle\delta y_a^2\rangle$, i.e., $\langle\delta y_a^2\rangle_{s}=\langle\delta y_a^2\rangle_{\rm min}=\kappa_{a}/(\kappa_{a}+4\Omega_{d})$~\cite{Scully97}, which decrease monotonically with the drive strength $\Omega_{d}/\kappa_{a}$, but the JPA becomes unstable for $\Omega_d/\kappa_{a} > 0.25$. Since the JPA can become unstable around the critical point $\Omega_d/\kappa_{a} = 0.25$, $\langle\delta y_a^2\rangle_{s}=\langle\delta y_a^2\rangle_{\rm min}=0.5$ is experimentally inaccessible. On the contrary, our proposed system is stable around the critical point and the optimum steady-state squeezing $\langle\delta y_a^2\rangle_{s}=0.5$ can be experimentally achievable [cf. Fig.~\ref{fig-squeezing-a}(b)]; note that Fig.~\ref{fig-squeezing-a}(b) is plotted by fixing the critical coupling strength $\chi_{c}$ (related to the drive strength $\Omega_{d}$) and varying the coupling strength $\chi$. When fixing $\chi$ and varying $\chi_{c}$, a similar figure can be obtained. Moreover, the proposed system is stable above the critical drive strength, where the minimum $\langle\delta y_a^2\rangle_{\rm min}$ is smaller than 0.5, which is also beyond the JPA [cf. Fig.~\ref{fig-squeezing-a}(b)].

\begin{figure}[tbp]
\centering
\includegraphics[width=0.48\textwidth]{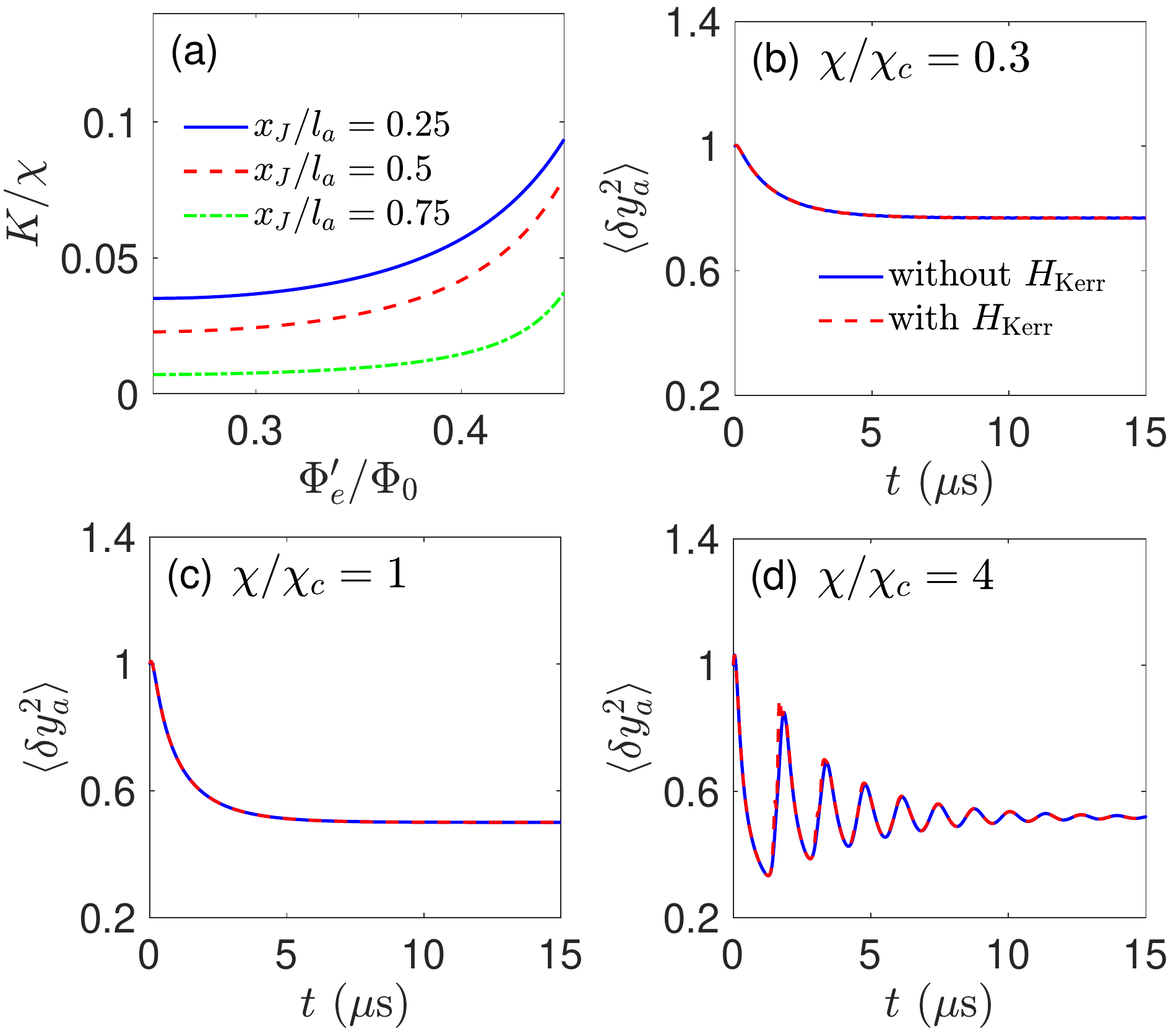}
\caption{(a) The ratio $K/\chi$ versus the external flux $\Phi'_{e}/\Phi_{0}$ for $x_{J}/l_{a}=$ 0.25, 0.5 and 0.75. Other parameters are the same as in Fig.~\ref{figure2}. (b)-(d) The time evolution of the variance $\langle\delta y_a^2\rangle$ in the two cases without and with the Kerr term, respectively, where the reduced coupling strength and the corresponding Kerr coefficient are (b) $\chi/\chi_c=0.3$, $K/\chi=0.008$, (c) $\chi/\chi_c=1$, $K/\chi=0.013$, and (d) $\chi/\chi_c=4$, $K/\chi=0.027$. Here the SQUID are located at $x_{J}/l_{a}=0.75$, and the different values of $\chi$ and $K$ in (b)-(d) correspond to different external flux $\Phi'_{e}$ [cf. green dashed-dotted curve in (a)]. Other parameters are the same as in Fig.~\ref{fig-s1}(a).}\label{fig-Kerr}
\end{figure}

\section{Discussions and conclusions}\label{conclusion}

In the above study, the higher-order nonlinear effect related to $\mathcal{L}_{S}^{(4)}$ is not considered. Below we show that this approximation is reasonable. With the relation in Eq.~(\ref{phip_m}),  we can obtain the Kerr Hamiltonian $H_{\rm Kerr}=-\mathcal{L}_{S}^{(4)}=-Ka^{\dag}aa^{\dag}a$ under the RWA, where the Kerr coefficient is
\begin{eqnarray}
K=\frac{1}{24L_{J}}\bigg(\frac{2\pi}{\Phi_{0}}\bigg)^{2}\bigg(\frac{\hbar\Delta_{1}^{2}}{2C_{\Sigma}{\omega_{a}}}\bigg)^{2}.
\end{eqnarray}
Figure~\ref{fig-Kerr}(a) displays the ratio of the Kerr coefficient $K$ to the coupling strength $\chi$ versus the external flux $\Phi'_{e}/\Phi_{0}$ at different positions $x_{J}/l_{a}$ of the SQUID. The corresponding $\chi$ versus $\Phi'_{e}/\Phi_{0}$ can be found in Fig.~\ref{figure2}(b). For different values of $x_{J}/l_{a}$, the ratio $K/\chi$ increases monotonically with $\Phi'_{e}/\Phi_{0}$. Also, for a fixed external flux $\Phi'_{e}/\Phi_{0}$, the ratio $K/\chi$ decreases when increasing $x_{J}/l_{a}$. Based on the results in Figs.~\ref{fig-Kerr}(a) and \ref{figure2}(b), it is vital to choose appropriate  $\Phi'_{e}/\Phi_{0}$ and $x_{J}/l_{a}$ to have a stronger coupling strength $\chi$ but still keep the ratio $K/\chi$ reasonably small.

Figures~\ref{fig-Kerr}(b)-\ref{fig-Kerr}(d) present the time evolution of the variance $\langle\delta y_a^2\rangle$ in the two cases without and with the Kerr term for different coupling strength $\chi/\chi_c$ (see Appendix~\ref{dynamic} for the detailed calculations). Obviously, the variance $\langle\delta y_a^2\rangle$ shows minor difference in these two cases for various values of $\chi/\chi_c$ (cf. the blue solid and red dashed curves). Therefore, it is reasonable to neglect the effect of the higher-order nonlinearities in our scheme. Moreover, only the microwave squeezing in CWR-A is investigated above. In fact, CWR-B can also exhibit the squeezing effect above the critical point (i.e., $\chi/\chi_{c}>1$), but weaker than CWR-A (see Appendix~\ref{squeezing-in-CWR-B}).

There are also other theoretical proposals for producing microwave squeezing using, e.g., superconducting resonant tank circuits~\cite{Zagoskin08}, circuit quantum electrodynamical systems~\cite{Benlloch14,Elliott15} and cavity magnomechanical systems~\cite{Li21}, but those schemes have not been explored experimentally. Different from those schemes, the JPA is widely studied in the experiment and our proposal provides an improved scheme modified from the flux-driven JPA. With the existing technologies, our proposed circuit can be easily fabricated in the experiment~\cite{Baust15}. Besides generating the squeezing, our scheme may have other potential applications in quantum technologies. For example, in our scheme, the nonlinear interaction of quantized-field-driven PDC can be achieved in the strong-coupling regime, because the typical decay rate of the CWR is smaller than $1\times2\pi$~MHz~\cite{Xiang13} and the strong-coupling condition $\chi>\{\kappa_{a},\kappa_{b}\}_{\rm max}$ can be well satisfied. With this strong-coupling quantized-field-driven PDC, the photon blockade becomes achievable~\cite{Zhou2020}, while it is inaccessible in the weak-coupling regime.

In conclusion, we have shown that the microwave squeezing can be considerably enhanced by designing a SQC, with two CWRs coupled via a SQUID embedded in one of the CWRs. In contrast to the flux-driven JPA~\cite{Zhong_2013}, optimum steady-state squeezing is experimentally accessible around the critical point and stronger transient squeezing is achievable above the critical drive strength in our scheme. Also, compared with the existing schemes of generating microwave squeezing via PDC~\cite{Moon05,Wang15}, our proposed SQC system can increase the nonlinear coupling strength by {\it three} orders of magnitude (i.e., from tens of kilohertz to tens of megahertz). This promises an appreciable enhancement of the microwave squeezing in the resonator.

\section*{Acknowledgments}

We acknowledge valuable discussions with Prof. Jian-Qiang You. This work is supported by the National Key Research and Development Program of China (Grant No.~2016YFA0301200), the National Natural Science Foundation of China (Grants No.~11774022, No.~U1801661, and No.~11934010), and Zhejiang Province Program for Science and Technology (Grant No.~2020C01019). G.-Q.Z. is supported by the Postdoctoral Science Foundation of China (Grant No. 2020M671687).

\appendix

\section{Normal modes of CWR-A}\label{normalmodes}

With the Lagrangian $\mathcal{L}'_{a}$ of bare CWR-A in Eq.~(\ref{bare}) and the linear term $\mathcal{L}_{S}^{(2)}$ of the Lagrangian of the SQUID in Eq.~(\ref{L_S1}), we can obtain the total linear Lagrangian $\mathcal{L}_{a}=\mathcal{L}'_{a}+\mathcal{L}_{S}^{(2)}$,
\begin{eqnarray}\label{La}
\mathcal{L}_{a}&=&\int_{-l_{a}}^{l_{a}}\bigg[\frac{C_{0}}{2}\dot{\Phi}^{2}(x,t)-\frac{[\partial_{x}\Phi(x,t)]^{2}}{2L_{0}}\bigg]dx \notag \\
                 &&+\frac{C_{J}}{2}\dot{\delta}^{2}- \frac{\delta^{2}}{2L_{J}}.
\end{eqnarray}
According to the Euler-Lagrange equation of motion, the flux field $\Phi(x,t)$ is found to obey the wave equation
\begin{eqnarray}\label{waveequation}
\ddot{\Phi}(x,t)-\upsilon^{2}\partial_{xx}\Phi(x,t)=0,
\end{eqnarray}
with the velocity $\upsilon=1/\sqrt{C_{0}L_{0}}$. By separating the variables, we write the solution in the form of
\begin{eqnarray}\label{phisolution}
\Phi(x,t)=\sum_{m}\phi_{m}(t)\mu_{m}(x),
\end{eqnarray}
where $\phi_{m}$ is the flux amplitude of the $m$th mode of CWR-A and $\mu_{m}(x)$ is the corresponding spatial mode function.
Substituting Eq.~(\ref{phisolution}) into Eq.~(\ref{waveequation}), we obtain two independent ordinary differential equations,
\begin{align}
~~~~~~\ddot{\phi}_{m}(t)+\omega_{m}^{2}\phi_{m}(t)&=0,\notag\\
\partial_{xx}\mu_{m}(x)+k_{m}^{2}\mu_{m}(x)&=0,
\end{align}
with the wave vector $k_{m}$ and the mode frequency $\omega_{m}=k_{m}\upsilon$.

To satisfy the boundary conditions in Eq.~(\ref{boundary1}) and the conditions in Eq.~(\ref{boundary2}) for the point $x=x_J$, the general solution of the spatial mode function $\mu_{m}(x)$ takes the form
\begin{eqnarray}\label{u_mx}
\mu_{m}(x)=A_{m}\bigg\{\begin{matrix}
           \sin[k_{m}(x+l_{a})-\varphi_{0}], &  -l_{a}\leq x \leq x_{J}^{-}, \\
           B_{m}\sin[k_{m}(x-l_{a})+\varphi_{0}], &  x_{J}^{+}\leq x \leq l_{a}.
                      \end{matrix}~~~~~
\end{eqnarray}

From Eq.~(\ref{boundary1}), the phase $\varphi_{0}=\pi/2$ can be obtained. By substituting the spatial mode function in Eq.~(\ref{u_mx}) into Eq.~(\ref{boundary2}), we get the expression for the relative amplitudes $B_{m}$,
\begin{eqnarray}
B_{m}=\frac{\cos[k_{m}(x_{J}+l_{a})-\varphi_{0}]}{\cos[k_{m}(x_{J}-l_{a})+\varphi_{0}]},
\end{eqnarray}
and the transcendental equation for the wave vectors $k_{m}$,
\begin{eqnarray}\label{transcendentalequation}
 &&\bigg(\tan[k_{m}(x_{J}-l_{a})+\varphi_{0}]-\tan[k_{m}(x_{J}+l_{a})-\varphi_{0}]\bigg)\notag\\
 &&\times\bigg[-(k_{m}l_{a})^{2}\frac{C_{J}}{C_{0}l_{a}}+\frac{L_{0}l_{a}}{L_{J}}\bigg] =k_{m}l_{a},
\end{eqnarray}
which can be numerically solved. The normalization constants $A_{m}$ can be fixed with the inner product relation~\cite{Goldstein80}
\begin{eqnarray}\label{innerrelation}
\langle\mu_{m}\cdot\mu_{n} \rangle &\equiv& \int_{-l_{a}}^{l_{a}}dx C_{0}\mu_{m}(x)\mu_{n}(x)+C_{J}\Delta_{m}\Delta_{n} \notag \\
                                   &=&C_{\Sigma}\delta_{mn},
\end{eqnarray}
where $\Delta_{m}=\mu_{m}(x_{J}^{+})-\mu_{m}(x_{J}^{-})$ is the mode-amplitude difference across the SQUID, and $C_{\Sigma}=2C_{0}l_{a}+C_{J}$ is the total capacitance. In addition, from Eqs.~(\ref{transcendentalequation}) and (\ref{innerrelation}), the inner product of the envelope derivatives are found to obey a similar orthonormality condition,
\begin{eqnarray}\label{um1}
\langle\partial_x\mu_{m}\cdot\partial_x\mu_{n} \rangle &\equiv& \int_{-l_{a}}^{l_{a}}dx \frac{1}{L_{0}}\partial_x\mu_{m}(x)\partial_x\mu_{n}(x)+\frac1{L_{J}}\Delta_{m}\Delta_{n} \notag \\
                           &=&\frac{\delta_{mn}}{L_m},
\end{eqnarray}
where $L_m$ defined by $1/L_m=C_{\Sigma}\omega_m^2$ is the effective inductance of the $m$th mode. Substituting the normal-mode expansion in Eq.~(\ref{phisolution}) into Eq.~(\ref{La}) and then performing spatial integration, we can write the Lagrangian $\mathcal{L}_{a}$ in the form of a set of harmonic oscillators,
\begin{eqnarray}
\mathcal{L}_{a}=\sum_{m}\left[\frac{1}{2}C_{\Sigma}\dot{\phi}_{m}^{2}-\frac{\phi_{m}^{2}}{2L_{m}}\right],
\end{eqnarray}
where the relations in Eqs.~(\ref{innerrelation}) and (\ref{um1}) are used.

\section{Dynamical equations of CWR-A and CWR-B}\label{dynamic}

When including the higher-order nonlinear effect of the SQUID, which corresponds to adding a Kerr term $-Ka^{\dag}aa^{\dag}a$ to the Hamiltonian $H$ in Eq.~(\ref{tot}), the total Hamiltonian of the system becomes
\begin{eqnarray}\label{B1}
H_{\rm tot}=-\chi(a^{\dag}a^{\dag}b+aab^{\dag})
           -Ka^{\dag}aa^{\dag}a+\Omega_d(b^{\dag}+b).
\end{eqnarray}

To study the microwave squeezing in the system, we define the variances $\langle\delta x_o^2\rangle\equiv\langle x_o^2\rangle-\langle x_o\rangle^2=\langle(\delta o^{\dag}+\delta o)^2\rangle$ and $\langle\delta y_o^2\rangle\equiv\langle y_o^2\rangle-\langle y_o\rangle^2=-\langle(\delta o^{\dag}-\delta o)^2\rangle$, with $o=a,~b$, where the Hermitian amplitude operators are given by $x_o=(o^{\dag}+o)$ and $y_{o}=i(o^{\dag}-o)$. Below we derive the variances $\langle\delta x_a^2\rangle$ and $\langle\delta y_a^2\rangle$ for the microwave field in CWR-A, as well as the variances $\langle\delta x_b^2\rangle$ and $\langle\delta y_b^2\rangle$ for the microwave field in CWR-B.

With the Hamiltonian in Eq.~(\ref{B1}), the dynamics of the proposed system is governed by the following Heisenberg-Langevin equations:
\begin{align}\label{HLab}
\dot a&=2i\chi a^{\dag}b+iK(aa^{\dag}a+a^{\dag}aa)-\frac{\kappa_{a}}{2}a+F_a(t),\notag\\
\dot b&=i\chi aa-i\Omega_{d}-\frac{\kappa_{b}}{2}b+F_b(t),
\end{align}
where $\kappa_{a}$ and $\kappa_{b}$ are the decay rates of the microwave fields in CWR-A and CWR-B, respectively. Here $F_a(t)$ and $F_b(t)$ are the corresponding input noise operators, which satisfy $\langle F_q(t)\rangle=0$, $\langle F_q(t)F_q(t')\rangle=\langle F_q^{\dag}(t)F_q^{\dag}(t')\rangle=0$,
$\langle F_q^{\dag}(t)F_q(t')\rangle=n_{q}\kappa_q\delta(t-t')$, and $\langle F_q(t)F_q^{\dag}(t')\rangle=(n_{q}+1)\kappa_q\delta(t-t')$, with $q=a,b$, where $n_{a}$ and $n_{b}$ are the numbers of thermal average photons in CWR-A and CWR-B. Due to the fact that the SQCs operate at milli-Kelvin temperatures~\cite{Gu17}, it is reasonable to ignore the temperature effect and take $n_{a}=n_{b}=0$. Following the procedures in Refs.~\cite{Wang15,Marshall90}, if we write the operators $a$ and $b$ as $a=\alpha+\delta a$ and $b=\beta+\delta b$, with $\alpha$ $(\beta)$ being the expectation value and $\delta a$ $(\delta b)$ the fluctuation, it follows from Eq.~(\ref{HLab}) that the average values $\alpha$ and $\beta$ satisfy
\begin{align}\label{alphabeta}
\dot \alpha&=2i\chi \alpha^*\beta+2iK|\alpha|^2\alpha-\frac{\kappa_{a}}{2}\alpha,\notag\\
\dot \beta&=i\chi \alpha^{2}-i\Omega_{d}-\frac{\kappa_{b}}{2}\beta,
\end{align}
and the fluctuation operators $\delta a$ and $\delta b$ obey
\begin{eqnarray}\label{deltaab}
\frac{d}{dt}\delta a&=&2i\chi \alpha^*\delta b+2(i\chi \beta+iK\alpha^2)\delta a^{\dag} \notag\\
                     &&+(4iK|\alpha|^2-\kappa_{a}/2)\delta a+F_a(t),\notag\\
\frac{d}{dt}\delta b&=&2i\chi \alpha\delta a-\frac{\kappa_{b}}{2}\delta b+F_b(t),
\end{eqnarray}
where the high-order terms of the fluctuations have been neglected. In the special case without the Kerr effect (i.e., $K=0$), the linearization procedure in Eq.~(\ref{deltaab}) gives the correct results both below and above the critical point, as discussed in Refs.~\cite{Wang15,Marshall90}. In our scheme, the Kerr coefficient is much weaker than the coupling strength [cf.~Fig.~\ref{fig-Kerr}(a) and related discussions], and the Kerr effect can be regarded as a perturbation~\cite{Boutin17}. Thus, the linearized approximation used in Eq.~(\ref{deltaab}) is well justified.

\begin{figure}[tbp]
\centering
\includegraphics[width=0.48\textwidth]{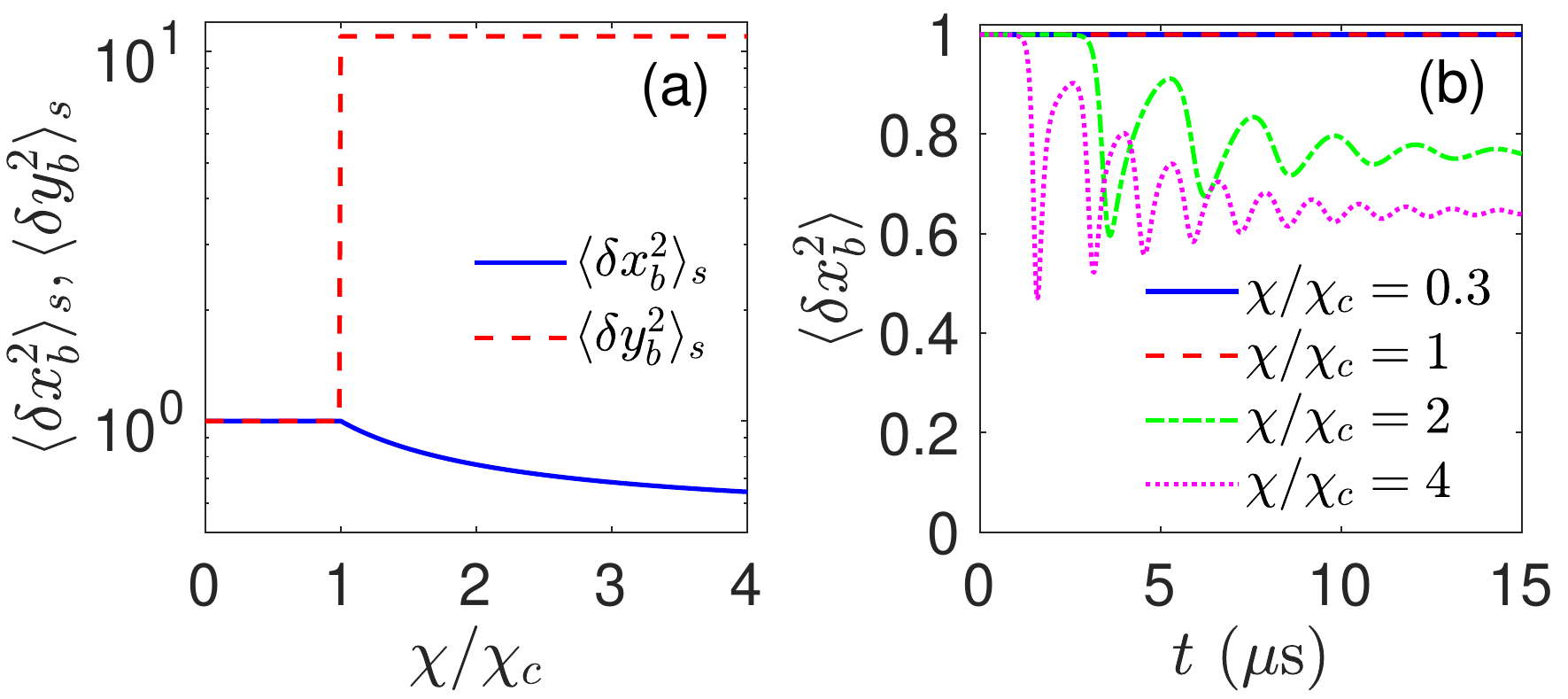}
\caption{(a) The steady-state variances $\langle\delta x_b^2\rangle_s$ and $\langle\delta y_b^2\rangle_s$ versus the reduced coupling strength $\chi/\chi_c$. (b) The time evolution of the variance $\langle\delta x_b^2\rangle$ for $\chi/\chi_c=0.3$, 1, 2 and 4. Other parameters are the same as in Fig.~\ref{fig-s1}(a).
}\label{fig-s2}
\end{figure}

To investigate the variances of the proposed system, we define the correlation parameters $A_1=\langle \delta a^2\rangle$, $A_2=\langle \delta a^{\dag}\delta a\rangle+\langle \delta a\delta a^{\dag}\rangle$, and $A_3=\langle \delta a^{\dag2}\rangle$ for CWR-A, which are related to the two-operator fluctuations. Using Eq.~(\ref{deltaab}), we obtain the equations of motion for $A_1$, $A_2$ and $A_3$,
\begin{eqnarray}\label{A1A2A31}
\dot A_1&=&4i\chi\alpha^*C_1+2(i\chi \beta+iK\alpha^2)A_2+2(4iK|\alpha|^2-\kappa_{a}/2)A_1 \notag \\
         &&+\langle\delta a F_a(t)+F_a(t)\delta a\rangle, \notag \\
\dot A_2&=&\left[4i\chi\alpha^*C_2+4(i\chi \beta+iK\alpha^2)A_3\right. \notag \\
         &&\left.+\langle\delta a^{\dag} F_a(t)+F_a(t)\delta a^{\dag}\rangle+h.c.\right]-\kappa_aA_2, \notag \\
\dot A_3&=&-4i\chi\alpha C_1^*-2(i\chi \beta^*+iK\alpha^{*2})A_2-2(4iK|\alpha|^2+\kappa_{a}/2)A_3 \notag \\
         &&+\langle\delta a^{\dag} F^{\dag}_a(t)+F^{\dag}_a(t)\delta a^{\dag}\rangle,
\end{eqnarray}
where $C_1=\langle\delta a\delta b\rangle$ and $C_2=\langle\delta a^{\dag}\delta b\rangle$ are the cross-correlation parameters for both CWR-A and CWR-B. Note that all correlation functions involving the noise operators in Eq.~(\ref{A1A2A31}) are zero, except for $\langle F_a(t)a^{\dag}\rangle=\langle aF_a^{\dag}(t)\rangle=\kappa_a/2$~\cite{Scully97}. Therefore, Eq.~(\ref{A1A2A31}) reduces to
\begin{align}\label{A1A2A3}
\dot A_1&=4i\chi\alpha^*C_1+2(i\chi \beta+iK\alpha^2)A_2+2(4iK|\alpha|^2-\kappa_{a}/2)A_1, \notag \\
\dot A_2&=[4i\chi\alpha^*C_2+4(i\chi \beta+iK\alpha^2)A_3+h.c.]-\kappa_aA_2+\kappa_a, \notag \\
\dot A_3&=-4i\chi\alpha C_1^*-2(i\chi \beta^*+iK\alpha^{*2})A_2-2(4iK|\alpha|^2+\kappa_{a}/2)A_3.
\end{align}
With similar procedures, the equations of motion for the cross-correlation parameters $C_1$ and $C_2$ are given by
\begin{eqnarray}\label{c1c2c3}
\dot C_1&=&2i\chi\alpha^*B_1+2i\chi\alpha A_1+2(i\chi \beta+iK\alpha^2)C_2 \notag \\
         &&+(4iK|\alpha|^2-\kappa_{a}/2-\kappa_{b}/2)C_1, \notag \\
\dot C_2&=&-i\chi\alpha(B_2-1)+i\chi\alpha(A_2-1)-2(i\chi \beta^*+iK\alpha^{*2})C_1 \notag \\
         &&-(4iK|\alpha|^2+\kappa_{a}/2+\kappa_{b}/2)C_2,
\end{eqnarray}
where the correlation parameters $B_1=\langle\delta b^2\rangle$, $B_2=\langle\delta b^{\dag}\delta b\rangle+\langle\delta b\delta b^{\dag}\rangle$ and $B_3=\langle\delta b^{\dag2}\rangle$ for CWR-B satisfy
\begin{align}\label{b1b2b3}
\dot B_1&=4i\chi\alpha C_1-\kappa_b B_1, \notag \\
\dot B_2&=4i\chi\alpha C_2^*-4i\chi\alpha^* C_2-\kappa_b B_2+\kappa_b, \notag \\
\dot B_3&=-4i\chi\alpha^* C_1^*-\kappa_b B_3.
\end{align}
Here we assume that the noise operators $F_a(t)$ and $F_b(t)$ are independent for the two resonators, i.e., $\langle F_a(t)F_b(t')\rangle=\langle F^{\dag}_a(t)F^{\dag}_b(t')\rangle=\langle F^{\dag}_a(t)F_b(t')\rangle=\langle F_a(t)F^{\dag}_b(t')\rangle=0$, under which $\langle aF_b(t)\rangle=\langle a^{\dag}F_b(t)\rangle=\langle bF_a(t)\rangle=\langle bF^{\dag}_a(t)\rangle=0$.
Obviously, the variances $\langle\delta x_a^2\rangle$, $\langle\delta y_a^2\rangle$, $\langle\delta x_b^2\rangle$ and $\langle\delta y_b^2\rangle$ can be expressed using the correlation parameters in Eqs.~(\ref{A1A2A3})-(\ref{b1b2b3}). With Eqs.~(\ref{alphabeta}) and (\ref{A1A2A3})-(\ref{b1b2b3}), we show the time evolution of the microwave-field amplitudes and the variances in Figs.~\ref{fig-s1}-\ref{fig-s2}, where the initial states in CWR-A and CWR-B are assumed to be coherent states $|\alpha_0=0.01\rangle$ and $|\beta_0=0.01i\rangle$, respectively, which correspond to $\alpha(t=0)=0.01$, $\beta(t=0)=0.01i$,  $A_1(t=0)=A_3(t=0)=B_1(t=0)=B_3(t=0)=C_1(t=0)=C_2(t=0)=0$, and $A_2(t=0)=B_2(t=0)=1$.

In addition to the dynamical behaviors, we are also interested in the steady-state behaviors of the proposed SQC system. Without the Kerr effect (i.e., $K=0$), Eq.~(\ref{alphabeta}) is reduced to Eq.~(\ref{steady}), and the corresponding steady-state amplitudes $\alpha_{s}$ and $\beta_{s}$, which satisfy $\alpha_{s}=\alpha_{s}^{*}$ and $\beta_{s}=-\beta_{s}^{*}$, are given in Eqs.~(\ref{solution-1}) and (\ref{solution-2}). Here we assume that the Rabi frequency $\Omega_{d}$ is real. Using Eqs.~(\ref{A1A2A3})-(\ref{b1b2b3}) and also with $\alpha=\alpha_{s}=\alpha_{s}^{*}$ and $\beta=\beta_{s}=-\beta_{s}^{*}$, we obtain the dynamical equations of the variances,
\begin{align}\label{dynamics-a}
\frac{d}{dt}\langle\delta x_a^2\rangle&=-4\chi\alpha_{s}\langle\delta x_a\delta y_b\rangle+(4i\chi\beta_{s}-\kappa_a)\langle\delta x_a^2\rangle+\kappa_a, \notag \\
\frac{d}{dt}\langle\delta x_a\delta y_b\rangle&=2\chi\alpha_{s}\langle\delta x_a^2-\delta y_b^2\rangle+\left(2i\chi\beta_{s}-\frac{\kappa_a+\kappa_b}{2}\right)\langle\delta x_a\delta y_b\rangle, \notag \\
\frac{d}{dt}\langle\delta y_b^2\rangle&=4\chi\alpha_{s}\langle\delta x_a\delta y_b\rangle-\kappa_b\langle\delta y_b^2\rangle+\kappa_b,
\end{align}
and
\begin{align}\label{dynamics-b}
\frac{d}{dt}\langle\delta y_a^2\rangle&=4\chi\alpha_{s}\langle\delta y_a\delta x_b\rangle-(4i\chi\beta_{s}+\kappa_a)\langle\delta y_a^2\rangle+\kappa_a, \notag \\
\frac{d}{dt}\langle\delta y_a\delta x_b\rangle&=-2\chi\alpha_{s}\langle\delta y_a^2-\delta x_b^2\rangle-\left(2i\chi\beta_{s}+\frac{\kappa_a+\kappa_b}{2}\right)\langle\delta y_a\delta x_b\rangle, \notag \\
\frac{d}{dt}\langle\delta x_b^2\rangle&=-4\chi\alpha_{s}\langle\delta y_a\delta x_b\rangle-\kappa_b\langle\delta x_b^2\rangle+\kappa_b.
\end{align}

Here Eqs.~(\ref{dynamics-a}) and (\ref{dynamics-b}) are valid in the limit $t\rightarrow +\infty$. At the steady states, the steady-state variances $\langle\delta x_a^2\rangle_s$, $\langle\delta y_a^2\rangle_s$, $\langle\delta x_b^2\rangle_s$ and $\langle\delta y_b^2\rangle_s$, as given in Eqs.~(\ref{variancex1}), (\ref{variancey1}), (\ref{variancex2}) and (\ref{variancey2}), can be derived by solving Eqs.~(\ref{dynamics-a}) and (\ref{dynamics-b}) with $d\langle\delta x_a^2\rangle/dt=d\langle\delta x_a\delta y_b\rangle/dt=d\langle\delta y_b^2\rangle/dt=d\langle\delta y_a^2\rangle/dt=d\langle\delta y_a\delta x_b\rangle/dt=d\langle\delta x_b^2\rangle/dt=0$.

\section{Microwave squeezing in CWR-B}\label{squeezing-in-CWR-B}

In Sec.~\ref{squeezing}, we have studied the microwave squeezing in CWR-A. Here we further show the microwave squeezing in CWR-B.
At the steady state, the variances $\langle\delta x_b^2\rangle$ and $\langle\delta y_b^2\rangle$ are obtained as (see Appendix~\ref{dynamic})
\begin{eqnarray}\label{variancex2}
\langle\delta x_b^2\rangle_s=\left\{\begin{matrix}
           1,  &  \chi<\chi_c; \\
           \dfrac{8\chi\Omega_d(\kappa_a+\kappa_b)+\kappa_a^2\kappa_b}{8\chi\Omega_d(2\kappa_a+\kappa_b)}, &  \chi>\chi_c,
                      \end{matrix}\right.
\end{eqnarray}
and
\begin{eqnarray}\label{variancey2}
\langle\delta y_b^2\rangle_s=\left\{\begin{matrix}
           1,  &  \chi<\chi_c; \\
           (\kappa_a+\kappa_b)/\kappa_b, &  \chi>\chi_c.
                      \end{matrix}\right.
\end{eqnarray}
When $\chi<\chi_{c}$ (i.e., below the critical point), no microwave squeezing occurs in CWR-B, because $\langle\delta x_b^2\rangle_s=\langle\delta y_b^2\rangle_s=1$. However, when the coupling strength $\chi$ is larger than the critical value $\chi_{c}$, $\langle\delta x_b^2\rangle_s<1$ and $\langle\delta y_b^2\rangle_s>1$ [cf. Fig.~\ref{fig-s2}(a)]. In this region, microwave squeezing can occur in CWR-B.

Compared with the steady-state squeezing in CWR-A, the squeezing effect in CWR-B is weaker (i.e., $\langle\delta x_b^2\rangle_s > \langle\delta y_a^2\rangle_s$) for a finite coupling strength $\chi$ [cf. Figs.~\ref{fig-s2}(a) and \ref{fig-squeezing-a}(b)]. In the limit $\chi/\chi_{c} \rightarrow +\infty$, the two resonators have the same squeezing effect with $\langle\delta y_a^2\rangle_s = \langle\delta x_b^2\rangle_s=0.524$. To further study the squeezing in CWR-B, we plot the time evolution of the variance $\langle\delta x_b^2\rangle$ for different values of the reduced coupling strength $\chi/\chi_c$ in Fig.~\ref{fig-s2}(b) (the related calculation details can be found in Appendix~\ref{dynamic}). Below the critical coupling strength, i.e., $\chi/\chi_{c}<1$, no microwave squeezing occurs in CWR-B (see the blue solid and red dashed curves). While above the critical coupling strength, i.e., $\chi/\chi_{c}>1$, microwave squeezing occurs in CWR-B (see the green dashed-dotted and magenta dotted  curves). By comparing Figs.~\ref{fig-s2}(b) and ~\ref{fig-squeezing-a}(c), we find that for a given coupling strength $\chi$, CWR-A exhibits a stronger squeezing effect than CWR-B.

\end{document}